\documentclass[%
 aip,
 amsmath,amssymb,
 reprint,%
]{revtex4-2}
\makeatletter

\usepackage{graphicx}
\usepackage{dcolumn}
\usepackage{bm}
\usepackage[dvipsnames]{xcolor}
\usepackage[utf8]{inputenc}
\usepackage[T1]{fontenc}
\usepackage{mathptmx}
\usepackage{etoolbox}
\usepackage{textcomp}
\usepackage{setspace}
\usepackage{graphicx}
\usepackage[english]{babel}
\usepackage{tabularx}
\usepackage{dcolumn}
\usepackage{bm}
\usepackage{tikz}
\usepackage{graphicx}
\usepackage{amssymb}
\usepackage{amsmath}
\usepackage{bm}
\usepackage{enumitem}
\usepackage{etoolbox}
\usepackage{wrapfig}
\def\@email#1#2{%
 \endgroup
 \patchcmd{\titleblock@produce}
  {\frontmatter@RRAPformat}
  {\frontmatter@RRAPformat{\produce@RRAP{*#1\href{mailto:#2}{#2}}}\frontmatter@RRAPformat}
  {}{}
}%
\makeatother
\begin{document}

\preprint{AIP/123-QED}
\title{Inter-particle adhesion induced strong mechanical memory in a dense granular suspension}
\author{Sebanti Chattopadhyay and Sayantan Majumdar*}
\email{smajumdar@rri.res.in}
\affiliation{Raman Research Institute, Bengaluru 560080, India}

\date{\today}

\begin{abstract}
Repeated/cyclic shearing can drive amorphous solids to a steady state encoding a memory of the applied strain amplitude. However, recent experiments find that the effect of such memory formation on mechanical properties of the bulk material is rather weak. Here we study the memory effect in a yield stress solid formed by a dense suspension of cornstarch particles in paraffin oil. Under cyclic shear, the system evolves towards a steady state showing training-induced strain stiffening and plasticity. A readout reveals that the system encodes a strong memory of the training amplitude ($\gamma_T$) as indicated by a large change in the differential shear modulus. We observe that memory can be encoded for a wide range of $\gamma_T$ both above and below the yielding, albeit, the strength of the memory decreases with increasing $\gamma_T$. In-situ boundary imaging shows strain localization close to the shearing boundaries, while the bulk of the sample moves like a solid plug. In the steady state, the average particle velocity $\left\langle v \right\rangle$ inside the solid-like region slows down with respect to the moving plate as $\gamma$ approaches $\gamma_T$, however, as the readout strain crosses $\gamma_T$, $\left\langle v \right\rangle$ suddenly increases. We demonstrate that inter-particle adhesive interaction is crucial for such strong memory effect. Interestingly, our system can also remember more than one input only if the training strain with smaller amplitude is applied last.
\end{abstract}

\maketitle

\section{Introduction}
Many out of equilibrium materials encode memory of past perturbations. The imprint of such perturbation history is stored in the material structure and can be revealed using a suitable readout protocol even long after the perturbation is removed. Cyclic perturbation protocols have been extensively used to encode memories in many of these systems. Such perturbations can be in the form of shear, temperature change, electrical/magnetic fields, etc. \cite{keim2019memory}. Few examples include dilute granular suspensions under cyclic shear \cite{Corte2008May, Keim2011Jun, Paulsen2014Aug}, aging and rejuvenation in glasses \cite{jonason1998memory, Zou2010Jun}, charge density wave conductors under voltage pulses \cite{fleming1986observation, coppersmith1997self}.
          
Memory formation in amorphous solids have attracted significant recent interests. Despite the diversity in microscopic details, presence of long-range correlations and complex energy landscapes, these  materials show very similar localized rearrangements under stress. Each of such rearrangements can be thought of as a transition between two local energy minima of the system \cite{falk2011deformation, cubuk2017structure}. Interestingly, these systems can also reach a steady state under cyclic deformations encoding a memory of the deformation amplitude. The approach to a steady state and memory formation in amorphous solids under cyclic shear has been demonstrated in numerical simulations of glassy and frictional granular systems \cite{Fiocco2013Aug, Fiocco2014Jan, Adhikari2018Sep, mungan2019networks, regev2013onset, keim2021multiperiodic, royer2015precisely}. Experimental studies have explored memory formation in soft glassy systems in both 2-D \cite{Mukherji2019Apr, keim2020global, keim2013yielding, keim2014mechanical} and 3-D \cite{nagamanasa2014experimental}, colloidal gels \cite{Schwen2020Apr}, cross-linked biopolymer networks \cite{schmoller2010cyclic, Majumdar2018Mar} . Many of these systems also have the ability to remember multiple inputs even in the absence of external noise.

The signature of the memory formed under cyclic shear is reflected as a sudden increase in particle mean square displacement (MSD) as the readout strain crosses the training strain amplitude marking an onset of irreversibility in the system. However, the effect of such reversible-irreversible transition on the mechanical properties of the bulk material has rarely been explored. Experiments on dilute non-Brownian suspensions \cite{Paulsen2014Aug} and a soft glassy system of 2-D bubble raft \cite{Mukherji2019Apr} reports that encoding memory induces only a small change in shear modulus of the system. These observations indicate a limitation of widely tuning the material properties using an imposed training.
 
In this Communication, we report strong memory formation in an amorphous solid formed by dense granular suspensions of cornstarch particles in paraffin oil. We find that memory can be encoded for a wide range of strain amplitudes both above and below yielding. Remarkably, such memory effect is directly reflected as a sharp change in the differential shear modulus of the system. We observe that in the case of consecutive training with the strain amplitudes $\gamma_1 \leq \gamma_2 \leq ...\leq \gamma_n$, only the memory of largest amplitude $\gamma_n$ is retained. However, if the system is trained with a smaller strain amplitude $\gamma_i (< \gamma_n$) at the end of the training sequence, then the memory of $\gamma_i$ can also be encoded. We show that such strong memory originates from a training-induced non-trivial coupling-decoupling dynamics of the solid-like region inside bulk of the sample with the shearing plate via high shear rate bands near the boundaries. We also demonstrate the crucial role played by the inter-particle adhesion in forming such strong mechanical memory.
\begin{figure}
    \begin{center}
    \includegraphics[height = 6 cm]{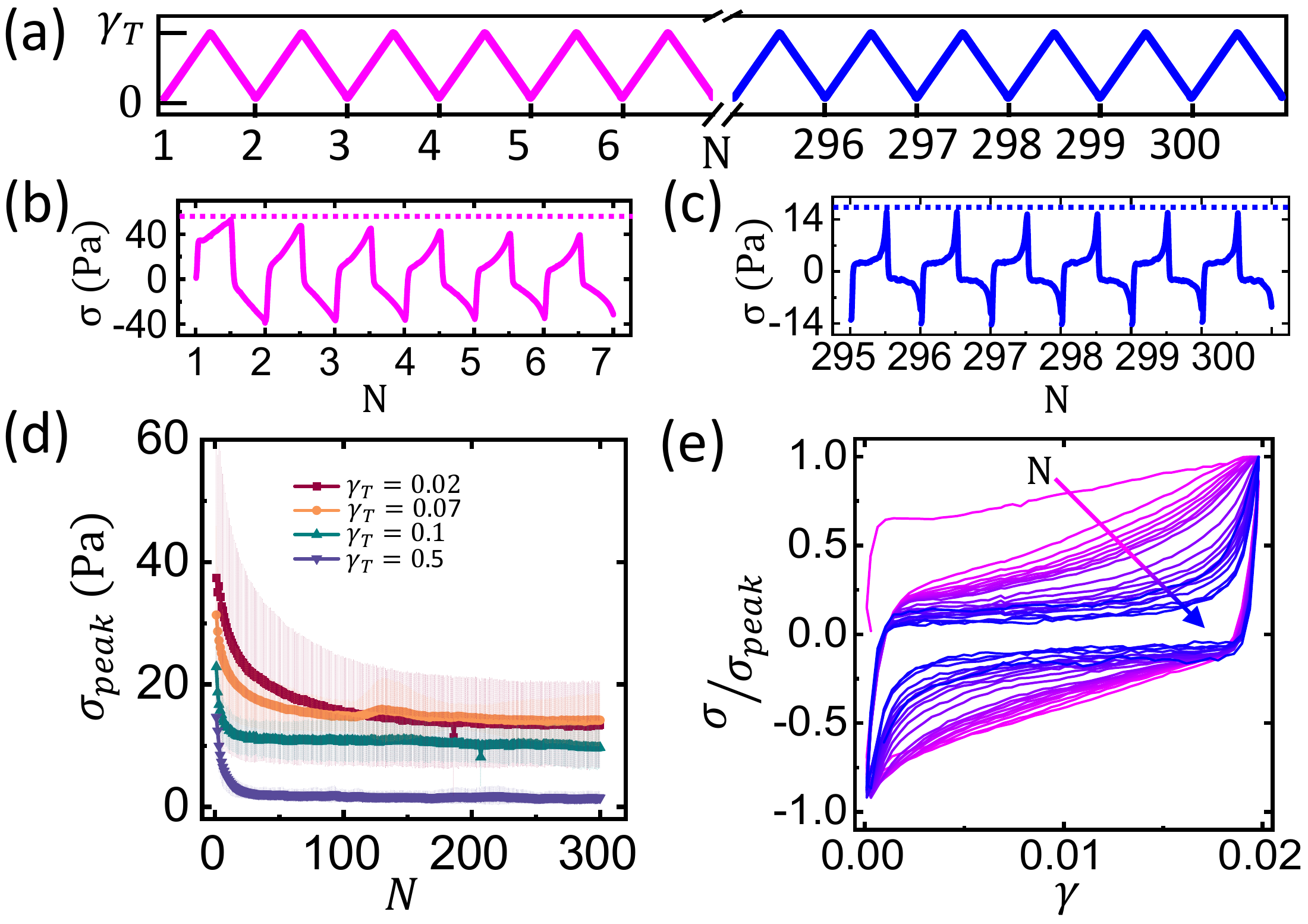}
    \caption{(a) Triangular wave strain deformation applied to the system for 300 cycles with a strain amplitude $\gamma_T$ as indicated in the figure. The magenta color indicates cycles 1-6 and the blue color indicates cycles 295-300 as shown. (b) and (c) show intra-cycle stress ($\sigma$) vs number of cycles (N) corresponding to the regions mentioned in (a) for $\gamma_T = 0.02$. The dotted lines correspond to the peak stress ($\sigma_{peak}$) for N = 1 (panel b) and N = 295 (panel c) showing that $\sigma_{peak}$ drops with increasing N in the beginning of the training, but finally saturates. (d) Variation of $\sigma_{peak}$ with N for different $\gamma_T$ values as indicated. The data is averaged over at least 3 independent measurements and error bars are the standard deviations. (e) Lissajous plots showing normalized intra-cycle stress ($\sigma /\sigma_{peak}$) vs. strain ($\gamma$) for few discrete N values. The arrow indicates increasing values of N.}
    \label{F1}
    \end{center}
\end{figure}
\section{Materials and methods}
For all our experiments, dense suspensions are prepared by dispersing Cornstarch (CS) particles (Sigma Aldrich) in paraffin oil with volume fraction ($\phi$) = 0.4. The CS particles have a mean diameter of 15$\pm 5 \mu$m. This system shows yield stress resulting from a solvent-mediated adhesion \cite{Richards2020Mar}. We use a MCR-702 stress controlled rheometer (Anton Paar) with a cone and plate geometry (C-P) having rough sand-blasted surfaces for all our measurements. In C-P, the shear-rate remains uniform everywhere inside the shearing geometry. 
The diameter of both cone and plate =  50 mm and the cone angle $\approx$ 2\textdegree. Our experiments are done in the separate motor transducer mode of the rheometer with the bottom plate moving and the cone remaining stationary at all times. We use an in-situ boundary imaging setup with a CCD camera (Lumenera) with a 5X long working distance objective (Mitutoyo) to capture the particle dynamics during the experiments. Images are captured at a rate of 4 Hz with a resolution of 1000 x 2000 pixels$^2$ for all the measurements. For varying the adhesive interaction in the system, we use the non-ionic surfactant Span\textsuperscript{\textcopyright}60. For more details about the system and the experimental set-up please see S.I. Notes and also \cite{chattopadhyay2022effect}. 
\begin{figure}
    \begin{center}
    \includegraphics[height = 4.8 cm]{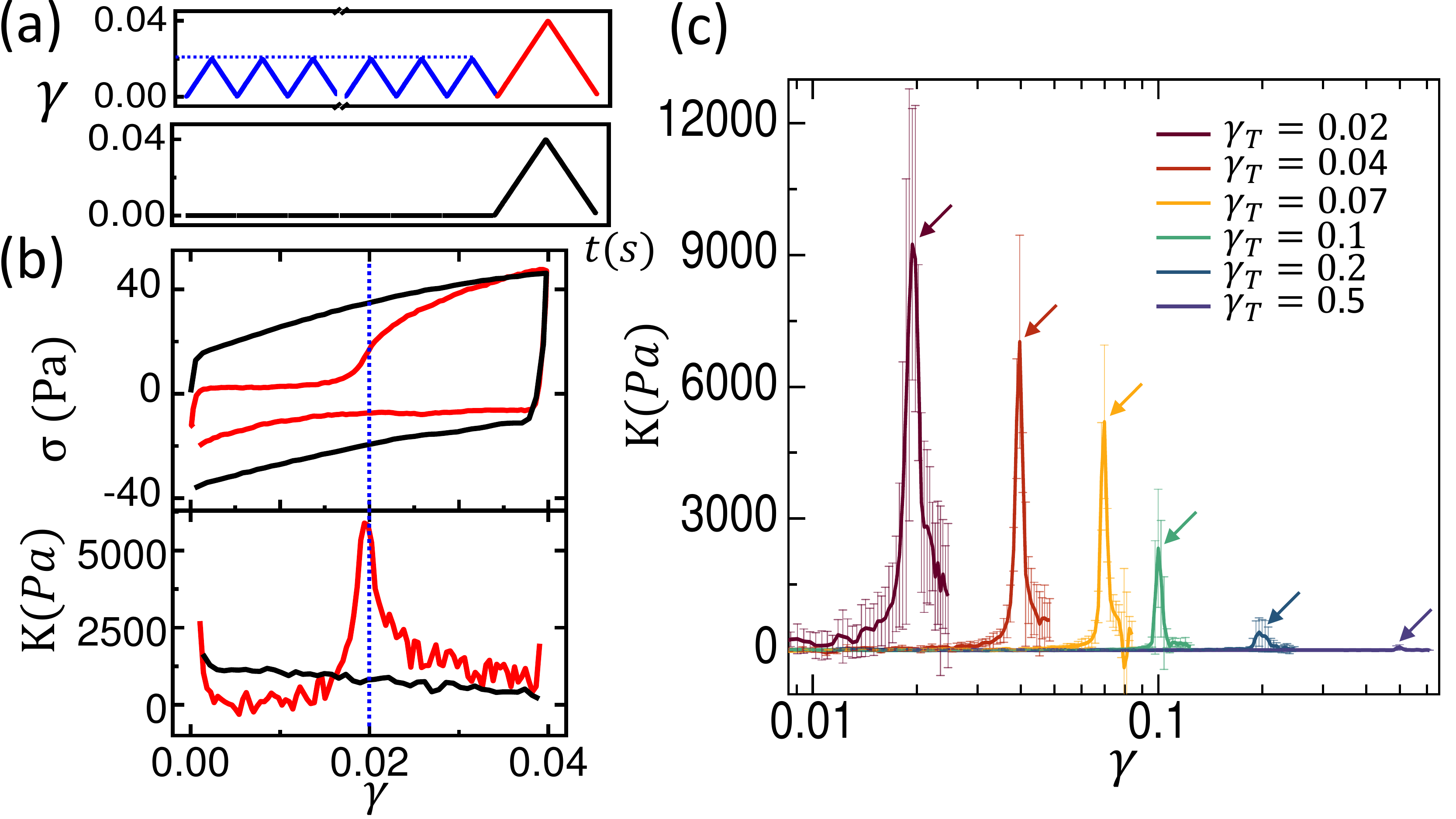}
    \caption{(a) Top: Training and readout protocol with $\gamma_{T}=0.02$ and $\gamma_{R}=0.04$, respectively. Bottom: Protocol of applying a single readout pulse ($\gamma_{R} = 0.04$) without any training. (b) Top panel: Stress ($\sigma$) vs strain($\gamma$) plots obtained from the readouts shown in (a) with training (in red) and without training (in black). Bottom panel: The corresponding differential shear moduli $K = \frac{d\sigma}{d\gamma}$ vs. $\gamma$ with same colour coding as the top panel. (c) Variation of $K$ vs. $\gamma$ for a wide range of $\gamma_T$ values as indicated in the legend with arrows indicating the position of the peak of $K$ for different $\gamma_T$ values. Each data set is averaged over at least 4 independent measurements. Error bars are the standard deviations.}
    \label{F2}
    \end{center}
\end{figure}
\begin{figure*}
    \begin{center}
    \includegraphics[height = 7 cm]{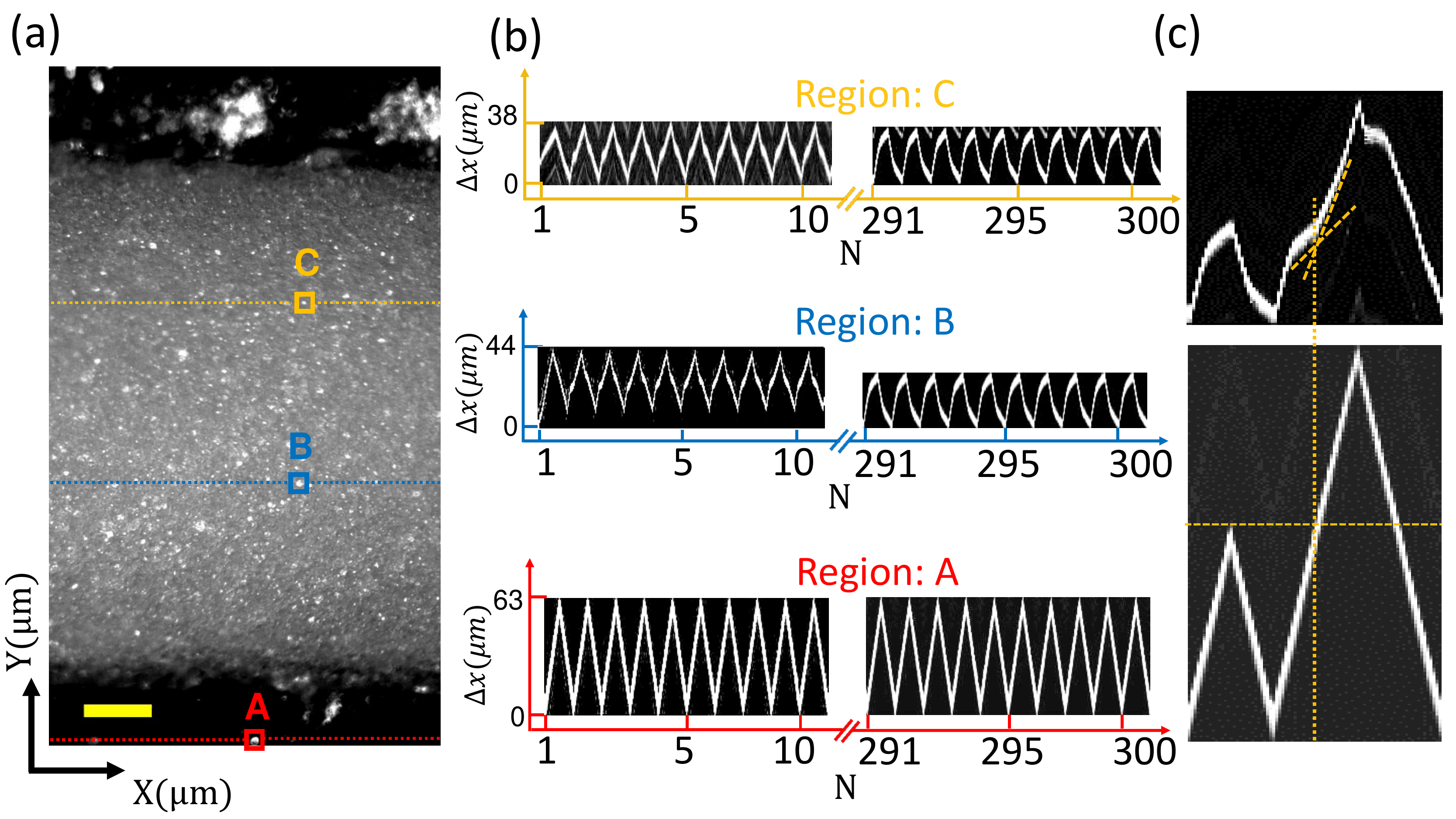}
    \caption{(a) A typical boundary image of the sample with the X-axis along the direction of flow and Y-axis along the gradient. Three points: A (on the plate), B and C (inside the bulk of the sample) indicated by boxes denote the location of the bright speckles for Kymograph analysis. The displacements of the bright speckles during the course of training are contained in the corresponding dotted lines. Corresponding Kymographs (for $\gamma_T$ = 0.07) are shown in panel (b) for the first 10 (N: 1 - 10) and last 10 (N: 291 - 300) training cycles, as indicated. (c) Kymographs corresponding to the last training (N = 300) and the readout cycle for the point A (bottom panel) and B (top panel). Dashed lines indicate a change in slope of the particle trajectory as the readout strain crosses $\gamma_T$. Scale bar in (a) denotes 100 $\mu$m. }
    \label{F3}
    \end{center}
\end{figure*}
\section{Results and discussion}
To train the system, we apply 300 cycles of a triangular-wave strain deformation with amplitude $\gamma_T$ at a constant strain rate of 0.01 $s^{-1}$, as shown in Fig. 1(a). At this shear rate, the Reynolds Number $Re << 1$, indicating that inertial effects can be neglected. We use different colors to indicate the cycles in the beginning (magenta) and in the end (blue). In Fig. 1(b) and (c), we show the resulting intra-cycle stress ($\sigma$) as a function of number of cycles (N) for $\gamma_{T}=0.02$. We notice that the peak stress ($\sigma_{peak}$) shows a systematic drop in the beginning (Fig. 1(b)) whereas near the end, it saturates to a steady state value of $\approx$ 15 Pa (Fig. 1(c)). Similar behaviour is also observed for other $\gamma_T$ values (Fig. 1(d)). From Fig. 1(d) we also see that $\sigma_{peak}$ for any N value decreases with increasing $\gamma_T$. This is related to the strain softening behaviour of the system: shear moduli $G'$ and $G''$ decrease with increasing strain amplitude as shown in Fig. S2. We note that the nature of the stress waveform also evolves with N. This is more clearly observed from the variation of normalized $\sigma$ vs. $\gamma$ (Lissajous plot) in Fig. 1(e). Out of total 300 cycles, we show the Lissajous plots for only a few discrete N values. We observe from Fig. 1(e) that starting from a quasi-linear visco-elastic response for small values of N, the system gradually shows a highly non-linear response for larger N values, approaching a steady state. Also, as we go to larger N values, the slope of the Lissajous plots remains negligible for $\gamma << \gamma_T$ however, near $\gamma_T$, the slope increases sharply. This indicates the development of plasticity and strain-stiffening behaviour under training. Similar behaviour has also been observed for colloidal gels and cross-linked biopolymer networks under cyclic shear \cite{vanDoorn2018May, schmoller2010cyclic}.  
\\\\Now, to see if the system encodes memory of the training amplitude, we apply a readout after 300 cycles of training. The readout is a triangular wave pulse having the same strain rate of 0.01$s^{-1}$ but with an amplitude $\gamma_R = 2\gamma_T$ as shown in Fig. 2(a) (top panel) for $\gamma_T = 0.02$. We plot $\sigma$ vs $\gamma$ of the readout cycle for $\gamma_{T} = 0.02$ in Fig. 2(b) (top panel, red curve). We observe a sharp change in the slope of $\sigma$ as we cross $\gamma = \gamma_T$. For a comparison, we apply the same readout to an untrained sample (Fig. 2(a), bottom panel) and plot the corresponding $\sigma$ vs. $\gamma$ in Fig. 2(b) (top panel, black curve) where we do not observe any such change. This difference is more clearly reflected in the differential shear modulus of the system $K (= \frac{d\sigma}{d\gamma}$) vs. $\gamma$ as shown in Fig. 2(b) (bottom panel): the trained sample shows a sharp peak in $K$ with value $K_{peak}$ at $\gamma = \gamma_T$, indicating that the system encodes a strong memory of the training amplitude which can change the differential shear modulus of the system by a huge amount. In our case $K_{peak} >> K_{baseline}$ (with $K_{baseline}$ as the value of $K$ at strain values slightly away from $\gamma_T$, where $K$ varies relatively slowly). Whereas, for earlier studies \cite{Paulsen2014Aug, Mukherji2019Apr} $K_{peak} \sim K_{baseline}$, further highlighting that memory formation is much stronger in our case. Additionally, we find that once the system is trained at a particular $\gamma_T$, any memory of $\gamma < \gamma_T$ gets erased (Fig. S3). We discuss the possibility of encoding multiple memories in our system later in the manuscript. We also find that the signature of encoded memory stays essentially the same, even when the readout is taken after 1000 s, implying that the relaxation of the structures supporting the memory is very slow. (not shown). 
\\To inspect the effect of training amplitude on the strength of memory formation, we encode and readout the memory for a range of $\gamma_T$ values. For each $\gamma_T$ we use a fresh loading of the sample. In Fig. 2(c), we plot average $K$ vs. $\gamma$ for various $\gamma_T$ values. We find a steady drop in the strength of the memory (quantified by the peak values of $K$) as $\gamma_T$ increases. In fact, beyond $\gamma_T$ = 0.1, $K_{peak}$ is significantly lower, albeit still present. This value of $\gamma_T$ is close to the fluidization/yielding onset of the system (Fig. S2). Thus, in contrast to 2-D soft glassy system \cite{Mukherji2019Apr}, we do not find any enhancement of memory effect near the yield point.  
\\\\So far we have established that the encoded memory gets reflected as a large change in the value of differential shear modulus of the system. In order to understand the mechanism behind this striking behaviour, we perform in-situ optical imaging (in the flow(X)-gradient(Y) plane). Details of the set-up are given in \cite{chattopadhyay2022effect}. A typical image is shown in Fig. 3(a) where the bright speckles correspond to CS particles which protrude out of the air-solvent interface. Internal connectivity of the fractal network giving rise to solid-like yield stress in the system stabilizes such protruding particles against the stress due to surface tension. We map out the velocity profile across the shear-gap using particle imaging velocimetry (PIV) during the training and readout experiments as shown in Fig. S4. We observe that except for narrow regions of high shear rate close to the shearing boundaries, the bulk of the sample moves like a solid plug with negligible velocity gradient. We use Kymographs to map out the space-time variation of the speckle distribution at different locations inside the sample acquired during training and readout experiments (Materials and Methods). We consider kymographs at three different locations: 1. on the moving plate (Region A), 2. inside the solid-like region close to the moving plate (Region B), and 3. inside the solid-like region close to the static cone (Region C). All these regions contain atleast one dominant bright speckle whose trajectory can be tracked during the training and readout. We show in Fig. 3(b) the kymographs corresponding to these regions for the first 10 (N = 1 - 10) and last 10 (N = 291-300) training cycles. We find that the trajectory obtained for Region A (Fig. 3(b), bottom panel) precisely mimics the motion of the shearing plate during training. The self-similarity of the waveform establishes the robustness of the input shear. Interestingly, Kymographs for Region 2 and 3 show more complex displacement waveforms involving a gradual evolution in both amplitude and shape before reaching a steady state during the course of training. However, except for the first cycle (involving start-up transients) waveform in these regions remains very similar, further confirming the solid plug-like motion of the bulk of the sample.
\\Interestingly, the displacement waveform inside the solid-like region of the sample indicates a complex coupling-decoupling dynamics between the moving plate and the bulk of the sample through the high shear rate bands near the boundaries. Clearly such dynamics is developed through training as it is absent in the beginning (left panels of Fig. 3(b)). In the steady state, particle trajectories in the solid-like region of the sample settle down to a waveform which is very different from the input, demonstrating a slowing down of the bulk as the intra-cycle input strain ($\gamma$) approaches the training amplitude ($\gamma_T$). Now, we take a look at the readout by plotting kymographs corresponding to the last training cycle (N = 300) and the readout cycle in Fig. 3(c). For clarity, we show the data only for Region A (bottom panel) and Region B (top panel) with expanded views. We find that the speckle-displacement during readout follows the last training cycle up to $\gamma = \gamma_T$. However, as $\gamma$ crosses $\gamma_T$, there is a sudden increase in the slope of the trajectory (marked by dashed lines in Fig. 3(c), top panel). Similar change in slope is also observed for other speckle trajectories inside the solid-like region (Fig. S1). This implies that beyond $\gamma_T$, there is a sudden buildup of coupling between the solid-like region and the moving plate. This sudden coupling will lead to an abrupt increase in velocity of a large number of particles, which can explain the origin of the sharp peak of $K$ around $\gamma = \gamma_T$. Such abrupt increase in velocity of the solid-like region inside the bulk of the sample also reflects in the velocity profile obtained from the PIV analysis (Fig. S5). Difference images constructed through stroboscopic sampling indicate a reversible to irreversible transition (RIT) in the sample beyond $\gamma_T$ (Fig. S6) similar to the earlier studies \cite{Paulsen2014Aug, Mukherji2019Apr, Fiocco2014Jan}.  
\begin{figure}
    \begin{center}
    \includegraphics[height = 4.9 cm]{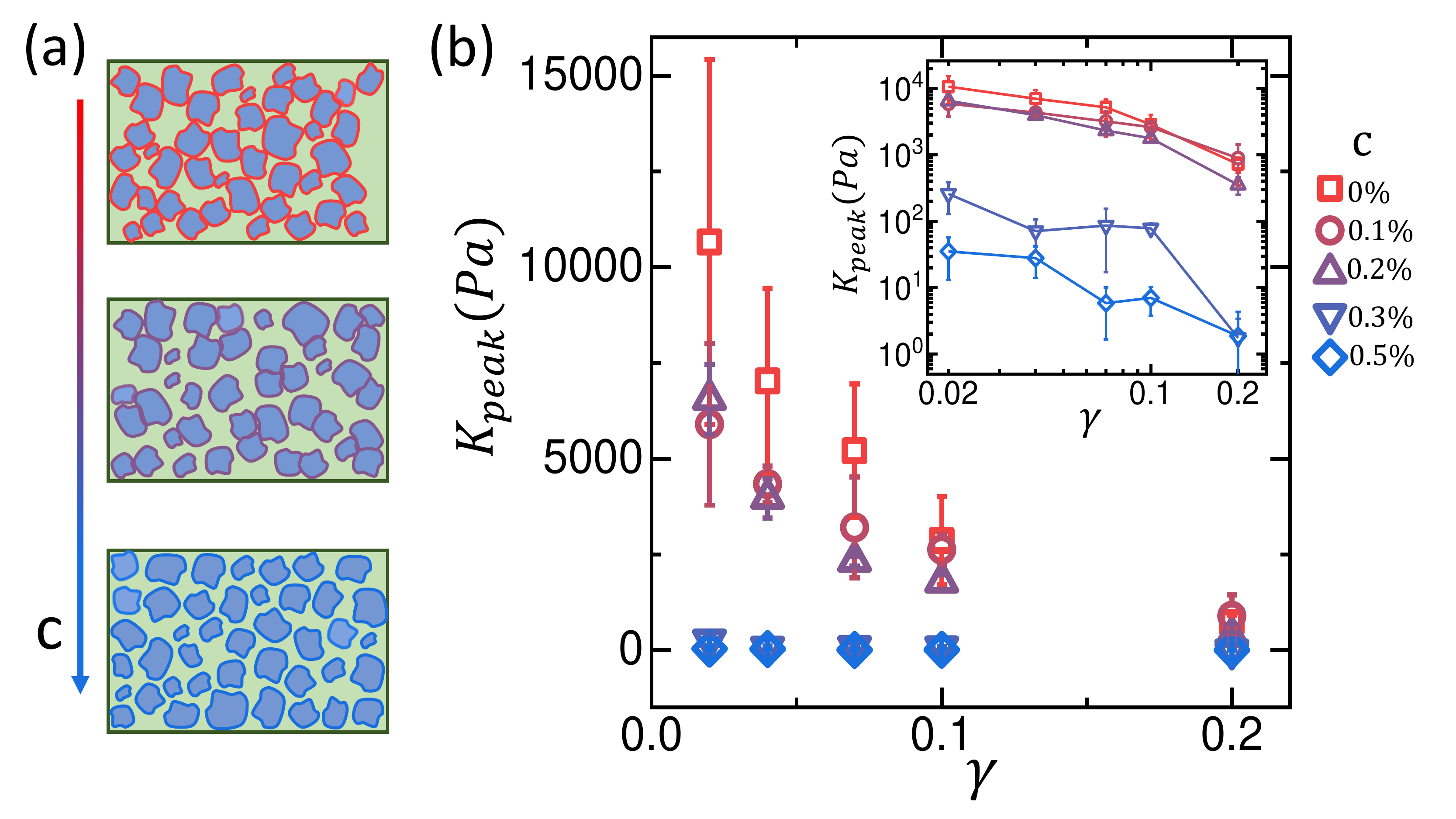}
    \caption{(a) Schematic indicating the gradual dispersion of fractal clusters with increasing surfactant concentration ($c$). For a given sample, $c$ denotes the mass fraction of surfactant w.r.t. corn starch particles. (b) Variation of the peak values of differential shear modulus $K_{peak}$ as a function of $\gamma_T$ for a range of $c$ values indicated. Each data set is averaged over at least 3 independent measurements with the error bars denoting the standard deviations. Inset shows a log-log plot of the same.}
    \label{F4}
    \end{center}
\end{figure}
\begin{figure}
    \begin{center}
   \includegraphics[height = 3.5 cm]{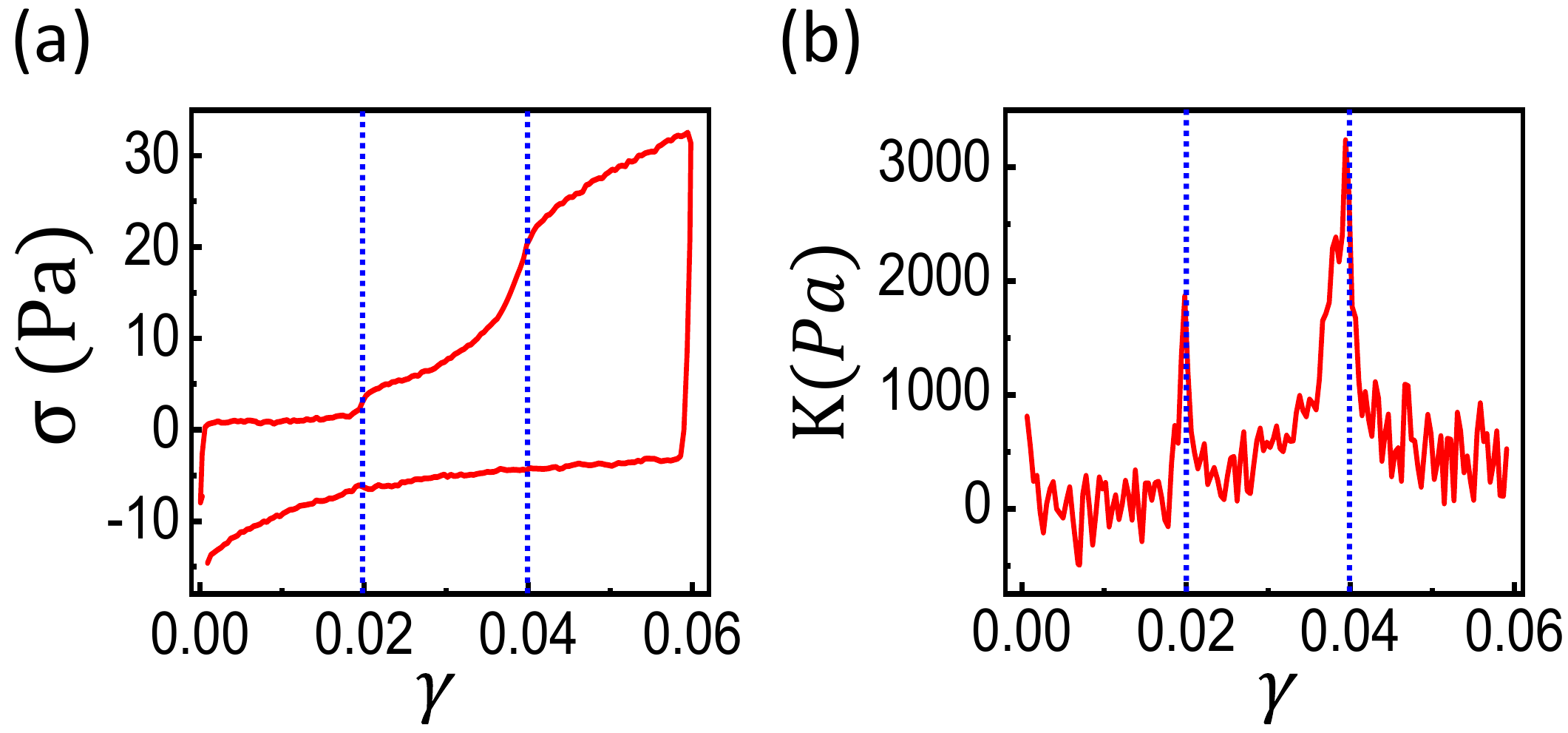}
   \caption{(a) Intra-cycle stress ($\sigma$) vs strain ($\gamma$) obtained from the readout for a sample trained at $\gamma_{T1}$ = 0.02 and $\gamma_{T2}$ = 0.04. (b) Corresponding $K$ vs $\gamma$ showing the signature of multiple memory formation.}
   \label{F5}
   \end{center}
\end{figure}   
\\\\Next, we address the role of inter-particle adhesive interaction on the strong memory formation in our system. Recently, we reported that addition of a small amount of surfactant to the cornstarch-paraffin oil system can reduce the adhesive interaction between the particles as quantified by the systematic change in jamming volume fraction \cite{chattopadhyay2022effect}. With sufficient amount of surfactant, the system completely transforms into a steric-repulsive one. Physically, as we increase the surfactant concentration ($c$), the large fractal clusters initially stabilized by inter-particle adhesion gradually disintegrate and finally form a well dispersed suspension as shown in the schematic Fig. 4(a). We prepare samples with increasing $c$ and subject them to the similar training and readout protocols mentioned earlier for different $\gamma_T$ values. In fig. 4(b), we plot the average peak values of $K$ for different $\gamma_T$ values obtained from the readouts for a range of $c$ values.  We see that the strength of memory ($K_{peak}$) decreases with increasing $c$ and the system loses ability to form strong memory beyond sufficiently high concentration of surfactant ($c \geq$ 0.3 \%). However, even at the highest surfactant concentration, the memory signature is not totally absent, as can be seen from the log-log plot (inset of Fig. 4b) and also Fig. S7.
We find that for the same range of $c$, the training induced strain stiffening also becomes significantly weaker (Fig. S8). Thus, the memory formation in our system is intimately related to the strain stiffening response of the system. From the Kymograph analysis we do not find any discontinuity in the particle trajectory across $\gamma_T$ in the readout for sufficiently high surfactant concentrations (Fig. S9).    
\\\\Lastly, we address the possibility of encoding multiple memories in our system. As mentioned earlier, if the system is trained at $\gamma_T$, it erases all memory of $\gamma < \gamma_T$. Interestingly, we find that if the sample is trained with the smaller strain amplitude just before the readout, it can retain the memory of the smaller $\gamma_T$ and more than one memory can be encoded, as has also been reported earlier for other systems \cite{Paulsen2014Aug, keim2020global}. We use the following protocol: we apply 1 cycle of the larger training strain amplitude ($\gamma_{T2}$ = 0.04) followed by 20 cycles of the smaller training strain amplitude ($\gamma_{T1}$ = 0.02). This whole sequence is repeated 15 times and then we apply a readout strain $\gamma_{R}$ = 0.06. The stress ($\sigma$) vs. strain ($\gamma$) plot for the readout shows sharp changes at $\gamma$ = 0.02 and 0.04 (Fig. 5(a)), giving two prominent peaks for the differential shear modulus $K$ (Fig. 5(b)). This indicates that the system has the ability to encode memories of more than one strain amplitude under a suitable training protocol.

\section{Conclusion}
We report strong memory formation in an adhesive dense particulate suspension under cyclic shear. The Differential shear modulus of the bulk system show a huge enhancement near the training strain amplitude ($\gamma_T$). A possible explanation for such striking effect is the following: in the presence of adhesion, the fractal particle clusters can form strand-like connected structures. Under cyclic shear deformations such structures can reorganize and develop a slack up to $\gamma_T$. The adhesive interaction maintains the contact between the particles in these loose strand-like structures for strain values $\gamma < \gamma_T$. Similar strand plasticity has been reported in colloidal gels under cyclic shear \cite{vanDoorn2018May}. For our case, such reorganization dynamics predominantly takes place in the high shear rate bands close to the shearing boundaries. Due to the slack in the particle strands, the plate can easily move without significantly disturbing the solid-like bulk region inside the sample for $\gamma < \gamma_T$. This gives rise to an apparent decoupling between the plate and the bulk sample in the steady state. However, when $\gamma$ crosses $\gamma_T$ during the readout, the loose strands suddenly get stretched and the coupling between the moving plate and the solid-like region builds up. This results in a strong stress response. Interestingly, such picture can also shed some light on the reduction in the strength of memory with increasing $\gamma_T$: for small $\gamma_T$ values, a large number of clusters can contribute to strand formation, however for large $\gamma_T$ values, many small clusters can get completely disintegrated and only large connected structures will contribute. Although, at present we cannot directly visualize the strand dynamics, the connection between the strand formation/plasticity and the observed memory effect is further supported by the role of adhesive interaction in the system. Once the inter-particle adhesion gets sufficiently weakened by addition of surfactant, the strand formation is no longer possible, as a result the memory effect gets diminished. This weaker memory in principle is similar to that observed earlier in dilute non-Brownian suspensions \cite{Paulsen2014Aug}. However, further work is needed to understand the exact correspondence. 

Memory effect reported here is reminiscent of similar phenomena in transiently cross-linked biopolymer networks \cite{schmoller2010cyclic} and Mullin's effect in filled-polymeric systems \cite{mullins1948effect}. Our system also shows a connection between the memory formation and reversible-irreversible transition like many repulsive particulate systems. Thus, adhesive particulate system can be thought of as an intermediate between repulsive dense suspensions and polymeric materials. Remarkably, our system presents a distinct advantage: since the particles are very robust and the adhesive particle-bonds are reversible, training and readout for a wide range of strain values does not cause any permanent damage to the system. The sample can be reloaded and reprogrammed arbitrarily. On the other hand, polymeric materials are prone to permanent damage due to breakage of filaments or chemical bonds. Interestingly, the strand picture together with the reversibility of adhesive bonds can explain the origin of multiple memories in our case. After encoding a memory at a smaller strain amplitude, application of a larger training strain can destroy the strands supporting the memory at smaller amplitude. However, after this, if the system is again trained at the smaller amplitude, the reversibility of the adhesive bonds ensures that some local connectivity can build up once more (Fig. S10), thus re-encoding the memory of smaller amplitude.

Due to the opaque nature of the particles, we can only probe the particle-dynamics on the sample surface. However, as our system is very dense (Fig. 3(a)), it is hard to directly probe training induced changes in the nodes or connectivity between the particles using boundary imaging alone. Mapping out the system dynamics in 3-D using optical/acoustic techniques remains a future challenge. Further theoretical insights regarding the mechanism of such strong memory effect in adhesive systems, including the formation of multiple memories can open up new avenues to explore. Our study can have important implications in designing programmable materials.
\section{Supplementary Material}
See supplementary material for information about sample preparation, data analysis and additional measurements.  
\section{Authors Contributions}
S.C. and S.M. designed the research, S.C. performed the experiments, S.C.and S.M. analyzed the data and wrote the manuscript.
\section{Conflict of interests}
There is no conflict to declare.
\section{Acknowledgments}
S.M. thanks SERB (under DST, Govt. of India) for a Ramanujan Fellowship. We acknowledge Ivo Peters for developing the Matlab codes used for PIV analysis. We thank Daniel Hexner for helpful discussion.

%
\clearpage

\begin{center}
\textbf{\Large{Supplementary Information}}
\end{center}
\vspace{1 cm}
\section{Supplementary note 1: Details of sample preparation and measurements}
When dispersed in oil, the cornstarch (CS) particles, which are hydrophilic due to the presence of -OH groups on their surface, experience a solvent-mediated adhesion. This adhesive interaction gives rise to a yield stress solid having system spanning fractal clusters for particle volume fraction $\phi > 0.23$. For all our measurements in this work, the particle volume fraction is $\phi=0.4$. To homogenize the samples, we use a combination of magnetic stirring and hand mixing. For making the samples with added surfactant, we dry mix the surfactant with the CS particles in a mortar pestle before adding oil to this mixture. We then homogenize the sample in a magnetic stirrer.
We perform triangular wave measurements as shown in Fig. 1 in the main text. We do not pre-shear the samples after loading as the pre-shearing can also encode some memory in the system. All our measurements are performed at a shear rate of 0.01 s$^{-1}$. We have also used higher shear rates for our experiments, but we have not found any systematic dependence of the strength of memory encoded on the shear rate over the range (0.01 - 0.1 s$^{-1}$) probed. Lower values shear rate lead to long experimental times, especially for larger training amplitudes, which may lead to sample drift. For this reason we have not explored shear rates lower than 0.01 s$^{-1}$.

\section{Supplementary note 2: Kymograph analysis}
We represent the space-time variation of surface intensity along a line segment using a Kymograph. For this, we first consider a series of images equally spaced in time. Then we choose a fixed line segment along the direction of shear (X-direction) (see Fig. 3 in the main text). The velocity gradient is along the Y-direction. Thus the segment represents a fixed value of y = $y_0$. The 2-D representation of the intensity distribution $I(x, y_0, t)$ of the line segment as a function of space and time gives the Kymograph. Interestingly, if a relatively bright speckle (representing a protruding particle at the air-liquid interface) moves along the chosen fixed line under shear and the line always contains the position of the speckle , then the Kymograph can directly represent the particle trajectory (Fig. S1).
\makeatletter
\newpage
\begin{figure*}[h!]
\renewcommand{\thefigure}{S1}
    \begin{center}
    \includegraphics[height = 9.8 cm]{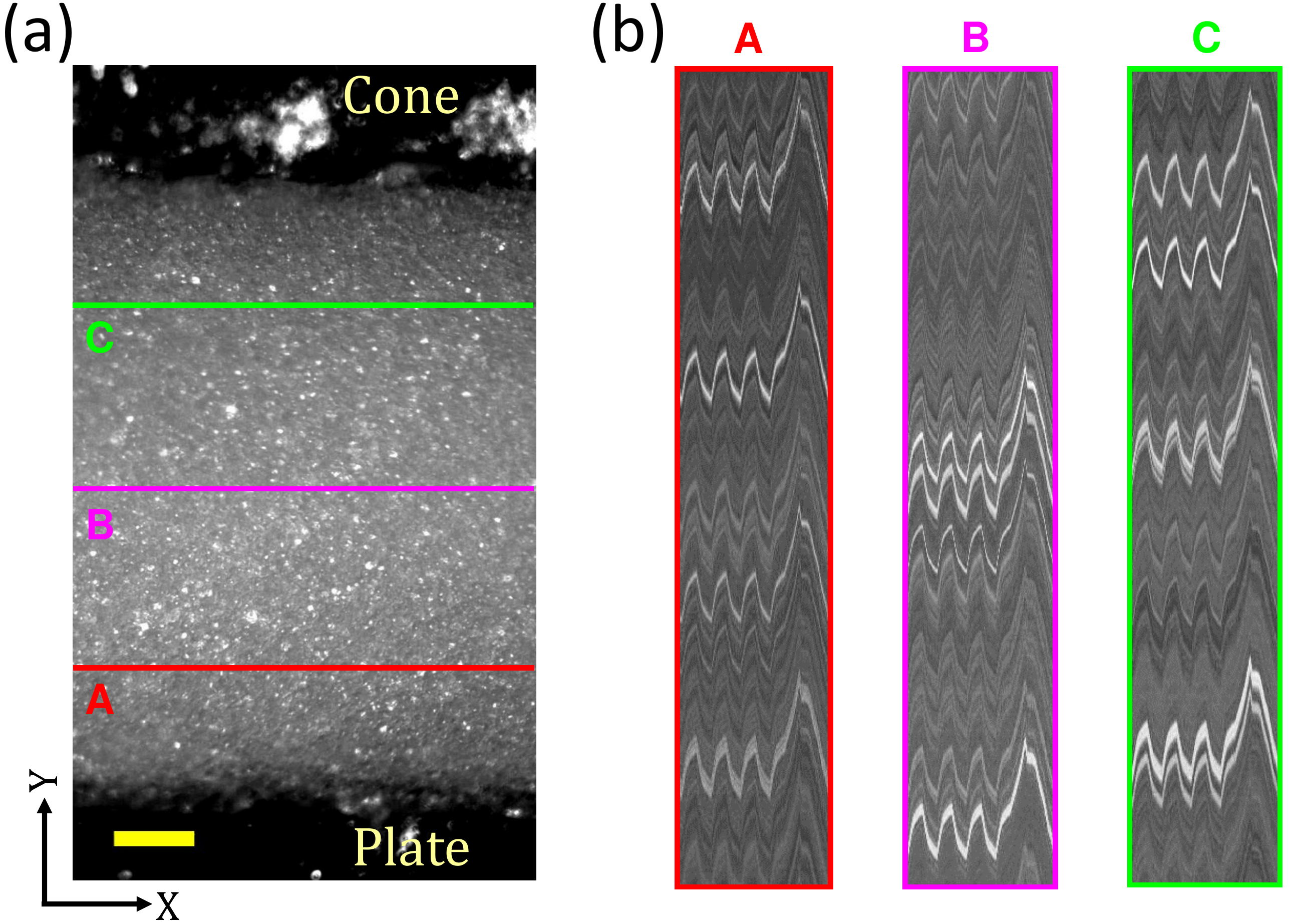}
    \caption{(a) A typical boundary image of the sample with lines depicting different regions inside the sample at varying distance from the moving plate. The scale bar represents 100 $\mu$m. (b) Corresponding Kymographs taken along the lines shown in (a). For clarity, only the last three training cycles (N: 298-300) and the readout is considered for each region. The multiple particle trajectories depicted in each Kymograph are the displacement waveforms for various speckles on the same line (see Materials and methods). The similarity of the waveforms, as well as, the discontinuity due to encoded memory indicates the solid plug-like motion of the bulk of the sample as also mentioned in the main text. Here $\gamma_T=0.07$.}
    \label{S1}
    \end{center}
\end{figure*}

\begin{figure*}
\renewcommand{\thefigure}{S2}
    \begin{center}
    \includegraphics[height = 7 cm]{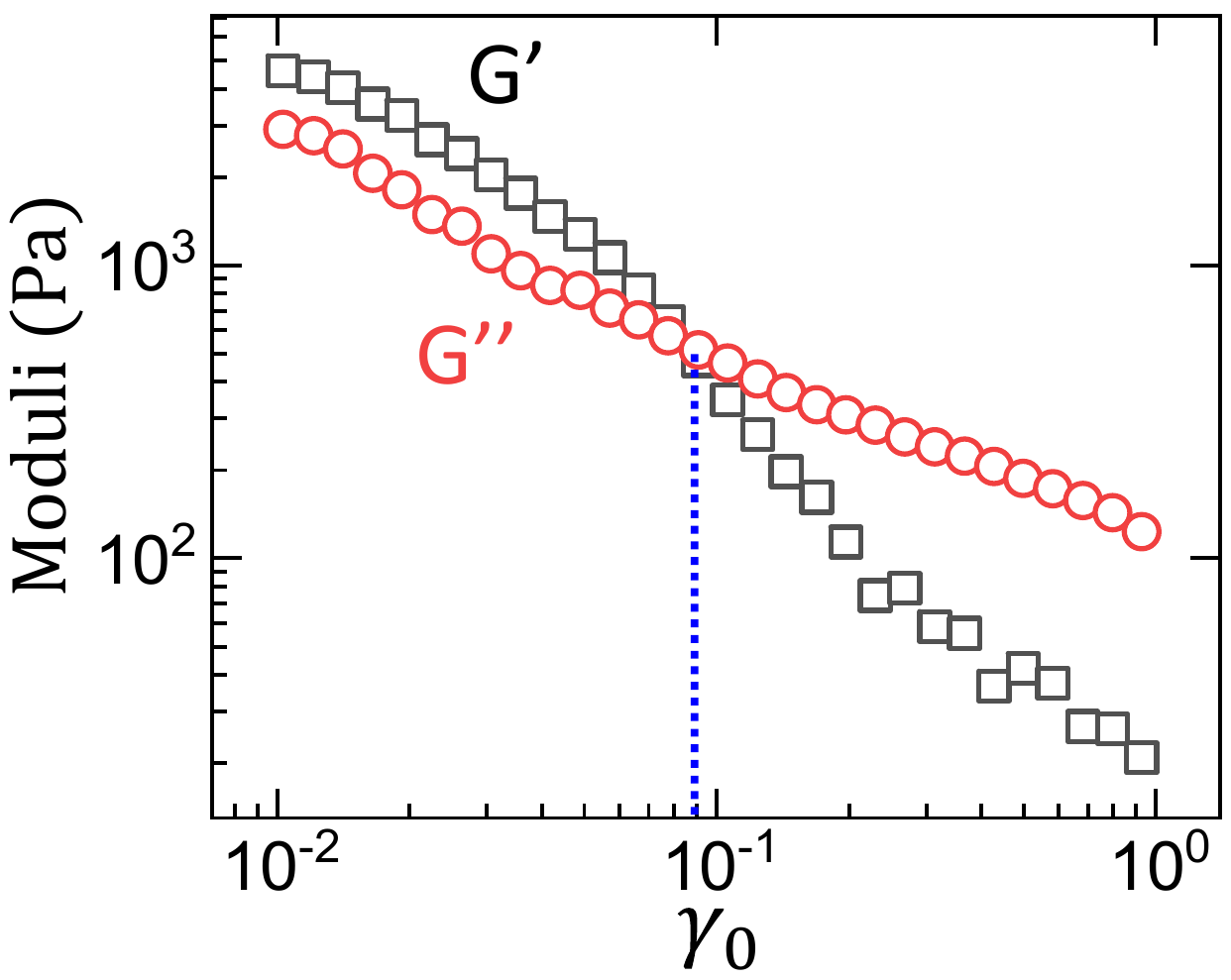}
    \caption{Variation of Elastic ($G'$) and Viscous ($G''$) moduli as a function of applied oscillatory strain amplitude ($\gamma_0$). The dashed line indicates the strain value for the crossover of $G'$ and $G''$ (yield strain), implying an onset of fluidization of the system. Here, we have used a sinusoidal waveform with angular frequency $\omega$ = 0.1 rad/s. We found that the value of yield strain remains independent of $\omega$ (we have checked over the range 0.05 to 0.5 rad/s). The yielding onset obtained from the normalized energy dissipation (see Ref.[27] in the main text) using triangular wave is in the same range as the yield strain obtained from the amplitude sweep data, as shown in this figure. Thus, in our case the nature of the imposed waveform does not affect yielding significantly (not shown).}
    \label{S2}
    \end{center}
\end{figure*}

\begin{figure*}
\renewcommand{\thefigure}{S3}
    \begin{center}
    \includegraphics[height = 6 cm]{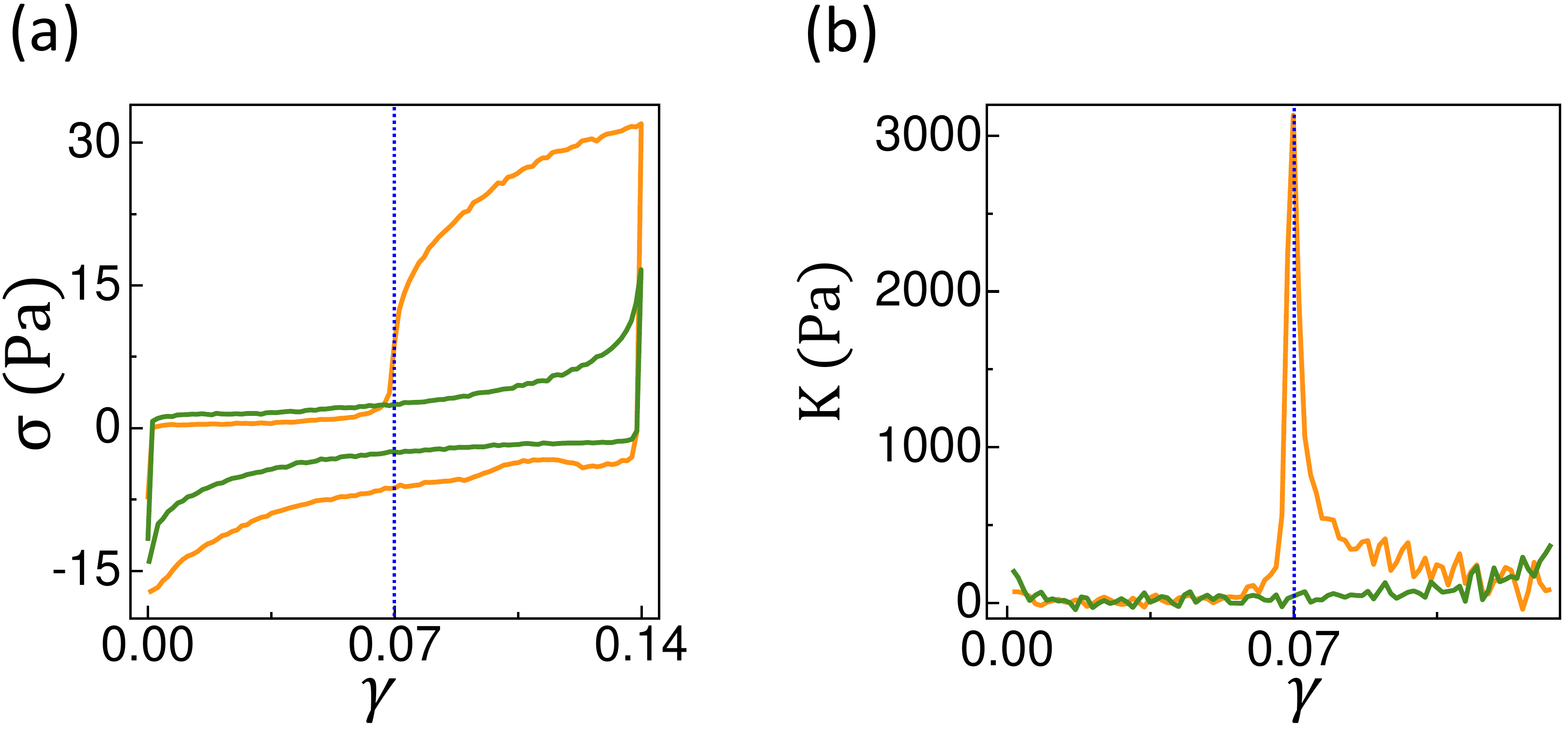}
    \caption{The system is first trained at $\gamma_T$ = 0.07. Immediately after this, the system is continued to be trained at $\gamma_T$ = 0.14 for a total of N = 20 cycles. We plot the intra-cycle stress ($\sigma$) vs. strain ($\gamma$) obtained for N = 1 (orange curve) and N = 20 (green curve) for $\gamma_T$ = 0.14 in (a). Corresponding $K$ vs. $\gamma$ plots are shown in (b). We see that while N = 1 shows a signature of memory at $\gamma_T$ = 0.07, there is no such signature for N = 20. Thus, the memory of the lower $\gamma_T$ = 0.07 gets completely erased within just 20 cycles of training at the larger strain $\gamma_T$ = 0.14. Dashed vertical lines indicate $\gamma$ = 0.07.}
    \label{S3}
    \end{center}
\end{figure*}

\begin{figure*}
\renewcommand{\thefigure}{S4}
    \begin{center}
    \includegraphics[height = 7 cm]{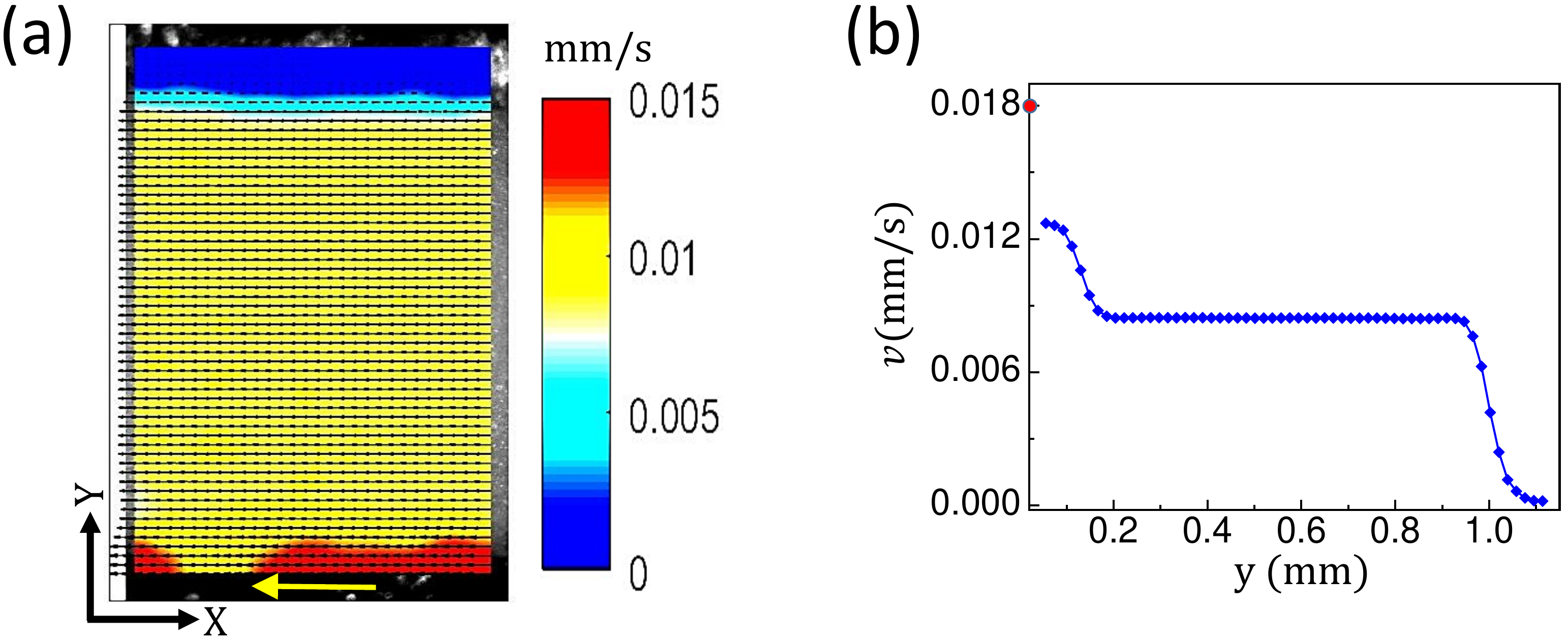}
    \caption{(a) A typical boundary image with superposed vector profile obtained from Particle Imaging Velocimetry (PIV) analysis (Materials and methods). (b) Velocity profile across the shear gap corresponding to the vector image shown in (a). Here, the velocity for each $y$ value represents the average over all the velocity vectors at the same $y$ coordinate but, varying $x$ coordinates.
		It is clear from the vector representation and the velocity profile that the system has high shear rate bands near the shearing boundaries, while the bulk of the sample is moving like a solid plug with almost a constant velocity. The red dot on the velocity axis indicates the calculated plate velocity.}
    \label{S4}
    \end{center}
\end{figure*}

\begin{figure*}
\renewcommand{\thefigure}{S5}
    \begin{center}
    \includegraphics[height = 6 cm]{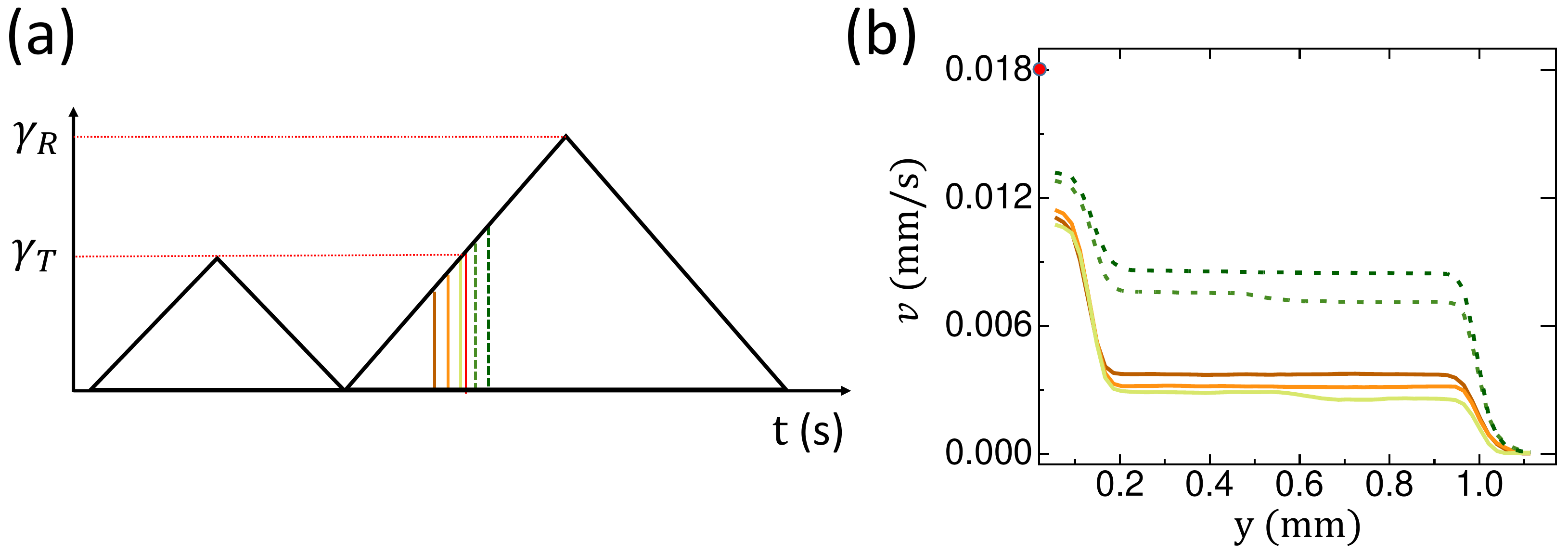}
    \caption{(a) A schematic showing the last training cycle (N = 300) and the readout cycle with dotted lines marking $\gamma_T$ and $\gamma_R$. Here $\gamma_T$ = 0.07 and $\gamma_R$ = 0.14. Vertical lines indicate different strain values (both above and below $\gamma_T$) during the readout for which the instantaneous velocity profiles are shown in panel (b). (b) Corresponding velocity profiles across the shear gap obtained through PIV analysis (see Fig. S4). Solid lines are profiles just below $\gamma_T$ and dashed lines are those just above $\gamma_T$. The colour coding is same as that in (a). We see that as the readout strain just crosses $\gamma_T$, the velocity of the solid plug-like region suddenly increases. This indicates a sudden buildup of contact between the solid plug and the moving plate. The red dot on the velocity axis indicates the calculated plate velocity.}
    \label{S5}
    \end{center}
\end{figure*}

\begin{figure*}
\renewcommand{\thefigure}{S6}
    \begin{center}
    \includegraphics[height = 10 cm]{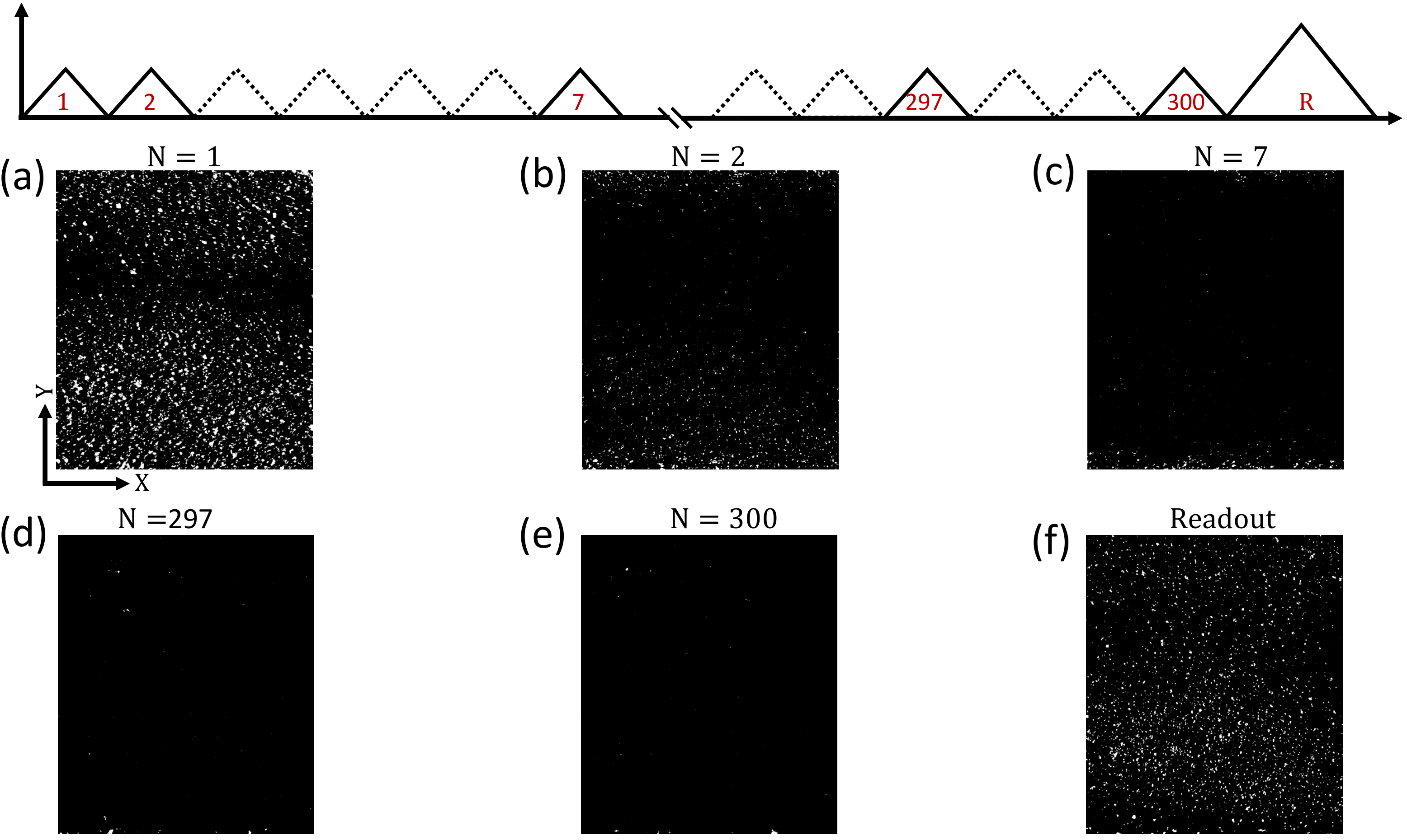}
    \caption{Stroboscopic difference between images of the system taken at the start and at the end of each cycle during the training and readout as shown in the schematic in the top panel. All images are cropped till the edges (including roughness) of the cone and plate. (a)-(e) Stroboscopic difference for different cycle number (N) indicated in the figure obtained during training. The bright spots in the difference images indicate the location of irreversible particle displacements that take place over the course of the N$^{th}$ cycle. We see that the particle rearrangements systematically drop with increasing number of training cycles and become almost negligible near the end of the training sequence indicating essentially reversible particle trajectories. Panel (f) represents the stroboscopic image corresponding to readout. The high intensity spots in (f) indicates the irreversibility of the particle trajectories. Clearly, a reversible to irreversible transition (RIT) happens as the system crosses $\gamma_T$. The data shown here is for $\gamma_T = 0.07$ and $\gamma_R$ = 0.14.} 
    \label{S6}
    \end{center}
\end{figure*}

\begin{figure*}
\renewcommand{\thefigure}{S7}
    \begin{center}
    \includegraphics[height = 5 cm]{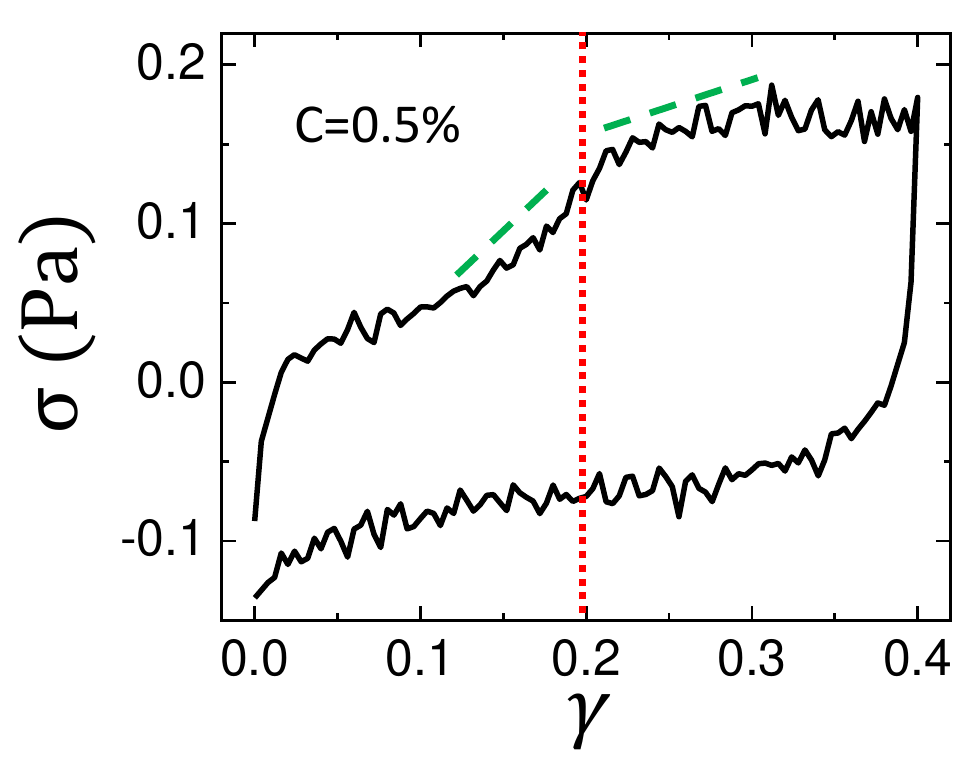}
     \caption{Readout Lissajous plot for $\phi=0.4$ with c = 0.5$\%$ as indicated in the figure. Dashed vertical line indicates the training strain amplitude. Green dashed lines along the curve indicate the change in slope across the training strain amplitude, showing that there is still a weak signature of memory even in the limit of negligible adhesion in the system. However, from the inset of main text figure 4(b), we see that the mean value of $K_{peak}$ is very small with large error bars. This is because, for low adhesion limit the absolute values of stress are quite low. This makes the the differential shear modulus data quite noisy.}
    \label{S7}
    \end{center}
\end{figure*}

\begin{figure*}
\renewcommand{\thefigure}{S8}
    \begin{center}
    \includegraphics[height = 12 cm]{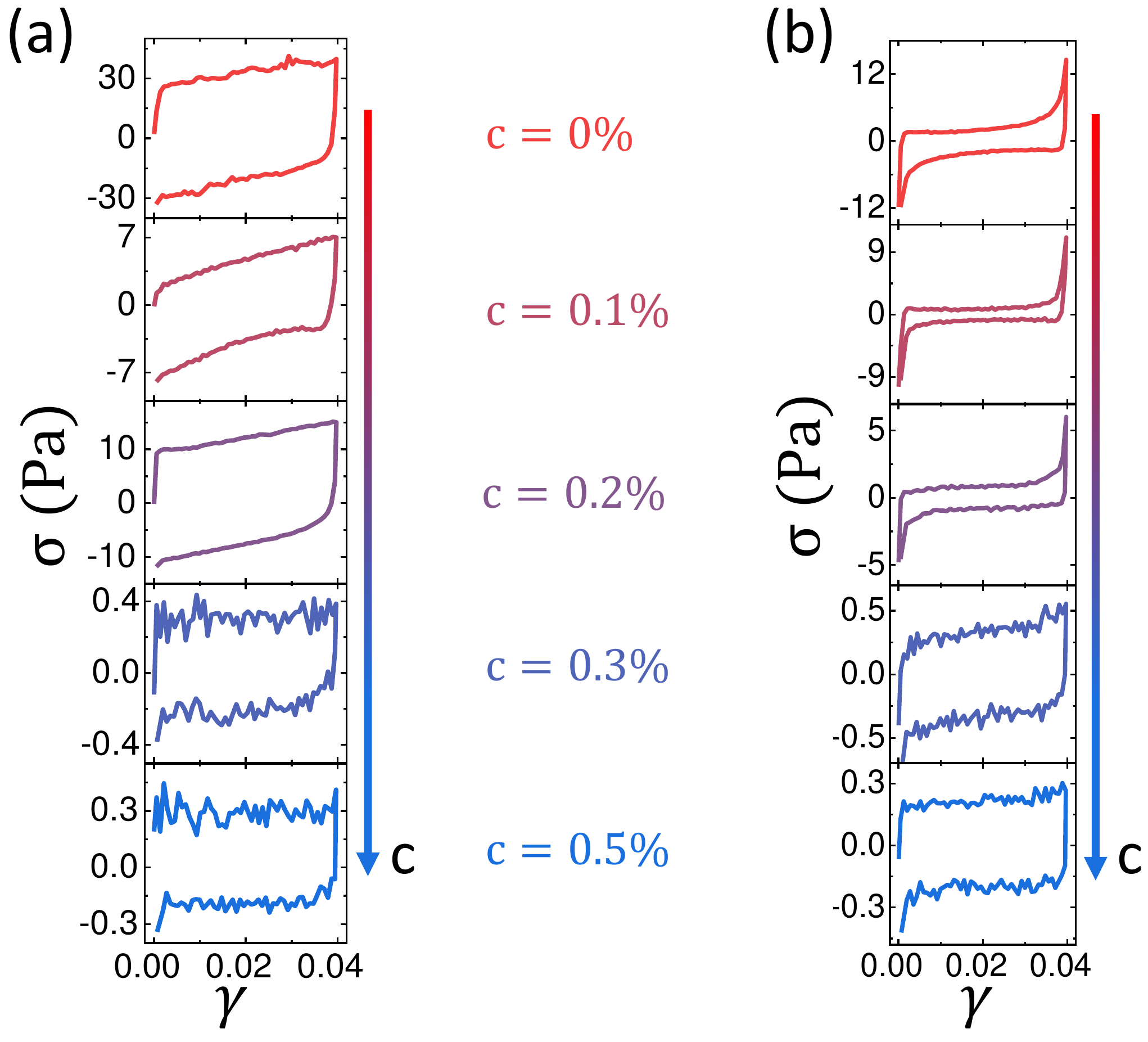}
    \caption{(a) Intra-cycle stress ($\sigma$) vs. strain ($\gamma$) corresponding to the first (N = 1, panel(a)) and the last training cycle (N = 300, panel (b)) for samples with different concentrations of surfactant ($c$) as indicated. The training induced strain stiffening behaviour gradually disappears with increasing surfactant concentration.}
    \label{S8}
    \end{center}
\end{figure*}

\begin{figure*}
\renewcommand{\thefigure}{S9}
    \begin{center}
    \includegraphics[height = 5 cm]{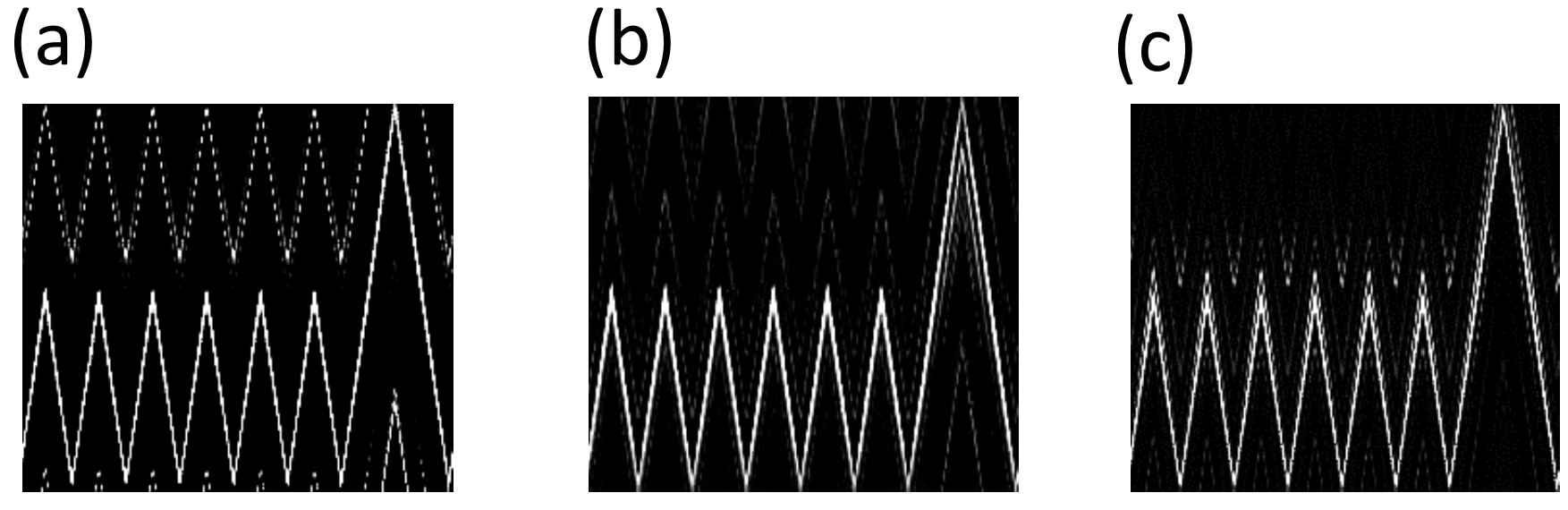}
    \caption{Kymographs of the last few training cycles (N: 295-300) and the readout cycle for the samples with a surfactant concentration $c = 0.5 \%$. Kymographs corresponding to a point on the moving plate (panel (a)), at a point inside the bulk but near the moving plate (panel (b)), at a point inside the bulk but near to the stationary cone (panel (c)). We see that there is no kink in the particle displacement waveform as the readout strain crosses the training strain amplitude (compare with Fig. 3(c) in the main text). This implies the absence of memory in the system for this concentration of surfactant. Here $\gamma_T$ = 0.07, and $\gamma_R$ = 0.14.}
    \label{S9}
    \end{center}
\end{figure*}
      
\begin{figure*}
\renewcommand{\thefigure}{S10}
    \begin{center}
    \includegraphics[height = 5 cm]{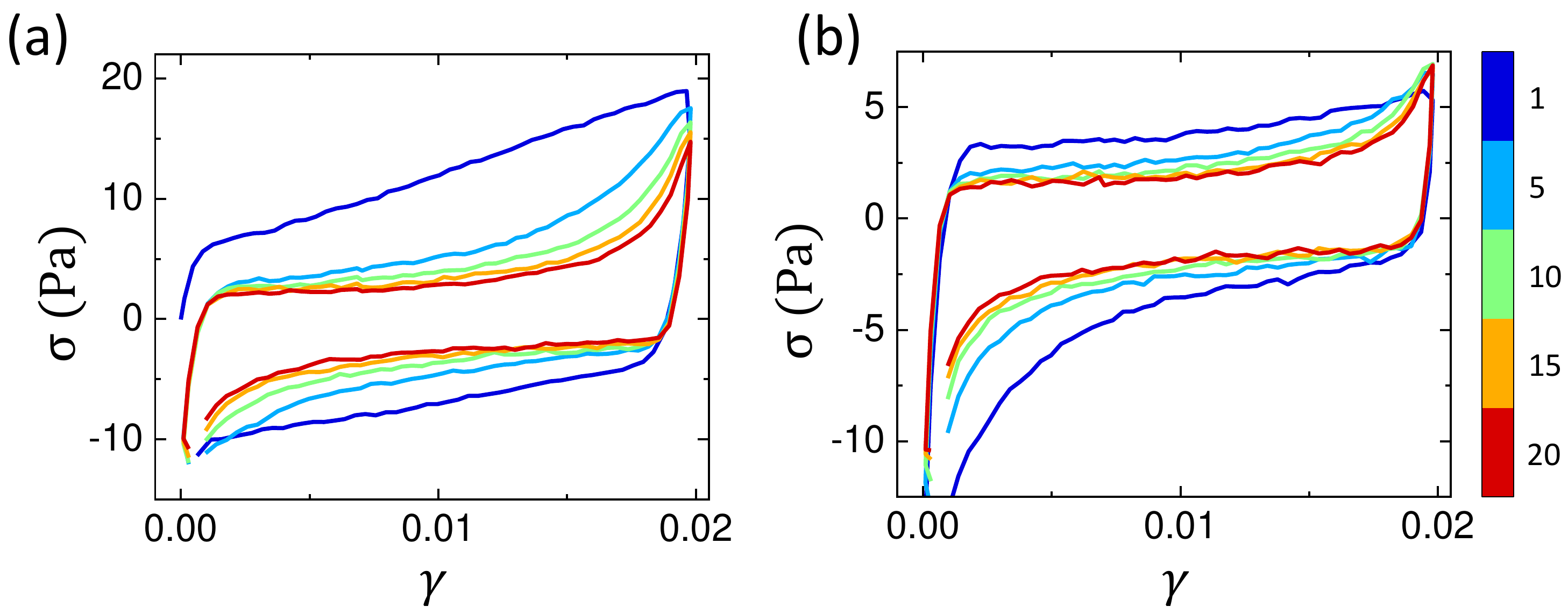}
     \caption{(a) The formation of strain stiffening with repeated application of strain amplitude $\gamma_{T1}$=0.02 for N=20 cycles. The curves are plotted at discrete values of N as indicated by the colorbar. (b) Following the 20 cycles shown in (a), we apply one cycle of $\gamma_{T2}$=0.04 (see main text for multiple memory protocol). After this, we apply the triangular wave with $\gamma_{T1}$=0.02 for 20 cycles once again. We see that while the initial curves show a very plastic behavior, by the 20$^{th}$ cycle, the system show some strain stiffened once again. On applying the larger amplitude of $\gamma_{T2}$=0.04, the encoded memory of $\gamma_{T1}$ gets destroyed as evidenced by the initial few curves of (b). However, on repeatedly applying $\gamma_{T1}$, the local structures are re-established and the system is able to strain stiffen at this amplitude once again, thus re-encoding the memory at $\gamma_{T1}$. This is possible because our system has inter-particle adhesive contacts that are physical bonds between the particles and thus, are reversible under breakage.}
    \label{S10}
    \end{center}
\end{figure*}

\end{document}